\documentclass[12pt]{article}
\newlength{\mytopmargin}
\newlength{\myleftmargin}
\usepackage{amsmath,amsthm,amssymb,amsbsy,epsfig,graphicx,color,multicol,subfigure}
\usepackage{array,calc}
\usepackage[enableskew]{youngtab}
\usepackage{rotating}
\usepackage{young,epic}
\usepackage{a4wide,bm}
\usepackage{url}
\usepackage{hyperref}

\newtheorem{prop}{Proposition}

\newtheorem{cor}{Corollary}

\numberwithin{equation}{section}

\usepackage{amsmath,amsfonts,amssymb}

\usepackage{graphicx}

\begin{document}
%
%\begin{frontmatter}

\title{Diffusion processes and the asymptotic  bulk gap probability for the real Ginibre ensemble}
\author{Peter J. Forrester }
\date{}
\maketitle
\noindent
\thanks{\small School of Mathematics and Statistics,  \\
 ARC Centre of Excellence for Mathematical \& Statistical Frontiers, \\
The University of Melbourne,
Victoria 3010, Australia;  email:  p.forrester@ms.unimelb.edu.au 
}

\begin{abstract}
It is known that the bulk scaling limit  of the real eigenvalues for the real Ginibre ensemble is equal in distribution to the
rescaled $t \to \infty$ limit of the annihilation process $A + A \to \emptyset$. Furthermore, deleting each particle at random
in the rescaled $t \to \infty$ limit of the coalescence process $A + A \to A$, a process equal in distribution to the annihilation
process results. We use these inter-relationships to deduce from the existing literature the asymptotic small and large
distance form of the gap probability for the real Ginibre ensemble. In particular, the leading form of the latter is shown
to be equal to $\exp(-(\zeta(3/2)/(2 \sqrt{2 \pi}))s)$, where $s$ denotes the gap size and $\zeta(z)$ denotes the Riemann zeta function. It is shown how this can be rigorously established using an asymptotic formula for matrix Fredholm operators. 
A determinant
formula is derived for the gap probability in the finite $N$ case, and this is used to illustrate the asymptotic formulas
against numerical computations.
\end{abstract}

\noindent 
{\it Dedicated to Professor R.J.~Baxter on the occasion of his 75${}^{\rm th}$ birthday.}

\section{Introduction}
For a point process on the line, a natural statistical quantity is the probability distribution for consecutive 
spacings between points. As an example, suppose the point process is a perfect gas in equilibrium with
density $\rho$. Then the probability density function for the distribution of consecutive spacings, $p_0(s)$ say,
is given by the Poisson distribution $p_0(s) = \rho e^{-\rho s}$, $s \ge 0$. A feature of the Poisson distribution
is that it is a maximum for zero separation, telling us that particles in a perfect gas tend to clump, while 
there is
exponential decay at large separation.

Our interest is in $p_0(s)$ for point process defined by the real eigenvalues of real standard Gaussian random matrices (matrices from
the real Ginibre ensemble \cite{Gi65}, to be abbreviated rG below). 
The interest in $p_0(s)$ for eigenvalues of random matrices goes back to Wigner in the 1950's (see
\cite{Po65} and references therein). The problem being addressed was that of providing a theoretical explanation for the statistical
properties of highly excited states of certain heavy nuclei, in sectors specified by definite quantum numbers. A significant property of the
energy levels is that they can be unfolded, which means scaled by the mean spacing to get parameter independent probability
distributions, and in particular a parameter independent form of $p_0(s)$. In relation to the latter Wigner hypothesized the functional form
\begin{equation}\label{1}
p_0^{\rm W}(s) = {\pi \over 2} s e^{- \pi s^2/4},
\end{equation}
now known as the Wigner surmise. Note that (\ref{1}) differs both in its small and large distance form to that of the Poisson distribution.
Specifically, 
the linear decay as $s \to 0$ quantifies the observed effect of repulsion between consecutive energy levels in experimental data,
while the exponential large $s$ form of the Poisson distribution is replaced by
a Gaussian.

In the course of this study Wigner introduced ensembles of real symmetric random matrices --- referred to as the Gaussian orthogonal ensemble
(GOE); see e.g.~\cite{Fo10}, as a theoretical model underlying (\ref{1}). Specifically, Wigner argued that the scaled bulk eigenvalues of
such matrices would have the same statistical distributions as those for the unfolded large energy levels of the complex nuclei. The task then was to
compute $p_0(s)$ for the bulk scaling limit of the GOE. In a major achievement, this was carried out by Gaudin \cite{Ga61}, who first expressed
$p_0(s)$ in terms of a Fredholm determinant, and then gave a computable expression for the latter in terms of prolate spherical functions.
(As an aside we remark that it is now realised that the Fredholm form determinant itself is well suited to direct numerical computation
\cite{Bo08,Bo09}.) This work demonstrated that (\ref{1}) is not exact, but nonetheless differs by at most a few percent in relative accuracy.
Later, building on the work \cite{JMMS80}, it was found \cite{FW00e} that the exact form of $p_0(s)$ in fact permits the general structure of (\ref{1}),
\begin{equation}\label{1a}
p_0(s) = {2 u((\pi s/2)^2) \over s} \exp \Big ( - \int_0^{(\pi s/2)^2} {u(t) \over t} \, dt \Big )
\end{equation}
((\ref{1}) corresponds to the choice $u(t) = t/\pi$), where $u$ is a particular solution of a certain $\sigma$ form Painlev\'e V nonlinear differential 
equation. The exact result (\ref{1a}) exhibits the small distance expansion
\begin{equation}\label{10.1}
p_0(s) = {\pi^2 s \over 6} - {\pi^4 s^3 \over 60} + {\pi^4 s^4 \over 270} + O(s^5),
\end{equation}
while for $s \to \infty$
\begin{equation}\label{10.2}
p_0(s) = e^{-(\pi s)^2/16 - \pi s/4 + (17/8) \log s + O(1)}
\end{equation}
(see e.g.~\cite{Fo12e} and references therein). Thus the small distance linear repulsion of (\ref{1}), and its large distance Gaussian decay are conserved by
the exact behaviours, although in both cases with different proportionality constants.

Our interest in this paper is to derive the analogue of the expansions (\ref{10.1}) and (\ref{10.2}) for the point process defined by bulk real eigenvalues
of the real Ginibre ensemble. For large matrix size $N$, it is known that the expected number of real eigenvalues is to leading order equal to
$\sqrt{2N/\pi}$, and the bulk density is equal to $1/\sqrt{2 \pi}$ \cite{Ed95}. This point process shares with the GOE eigenvalues the special feature that it
is an example of a Pfaffian point process. Thus the $k$-point correlation function can be expressed as a $2k \times 2k$ Pfaffian with matrix elements
independent of $k$. In the bulk scaling limit this reads \cite{FN07,Som07,BS09}
\begin{equation}\label{8.2}
\rho_{(n)}^{\rm rG}(\lambda_1,\dots,\lambda_n) = {\rm Pf}
\begin{bmatrix} - {I}(\lambda_i,\lambda_j) & S(\lambda_i,\lambda_j) \\
- S(\lambda_j,\lambda_i) & D(\lambda_i,\lambda_j) \end{bmatrix}
\end{equation}
where
\begin{align}\label{8.2a}
S(x,y) & = {1 \over \sqrt{2 \pi}} e^{-(x-y)^2/2} \nonumber \\
{I}(x,y) & = {1 \over 2} {\rm sgn}(y-x) - \int_{x}^y S(x,u) \, du  \nonumber \\
D(x,y) & = {\partial \over \partial x} S(x,y).
\end{align}
Recently it has been proved that (\ref{8.2}) and (\ref{8.2a}) hold for the real eigenvalues of a much larger class of non-Hermitian
random matrices with real entries than just the real Ginibre ensemble \cite{TV12}.

In general the $k$-point correlation functions can be used to compute spacing distributions. Thus let $E(k;J)$ denote the probability that the
interval $J$ contains exactly $k$ eigenvalues (in particular, $E(0;(0,s))$ is referred to as the gap probability), and define the
corresponding generating function by
$$
E(J;\xi) = \sum_{k=0}^\infty (1 - \xi)^k E(k;J).
$$
Then we have
\begin{equation}\label{8.3}
E(J;\xi) = 1 + \sum_{n=1}^\infty {(-\xi)^n \over n!} \int_J dx_1 \cdots \int_J dx_n \, \rho_{(n)}(x_1,\dots,x_n).
\end{equation}
It is of interest to remark that $E(J;\xi)$, for $0 < \xi \le  1$, also has the interpretation of there being no eigenvalues (and thus
as a gap probability) in the interval $J$ of the original system diluted so that each eigenvalue
is independently deleted with probability $(1 - \xi)$ (see e.g.~\cite[paragraph below (9.4)]{Fo10}). In the special case
that $\rho_{(n)}$ has the Pfaffian structure (\ref{8.2}), it is generally true that (\ref{8.3}) can be summed to give a Fredholm
determinant. In this regards, let $Z_2 := \begin{bmatrix}0 & -1 \\ 1 & 0 \end{bmatrix}$, and transform the $2 \times 2$ blocks in
(\ref{8.2}) by multiplication on the right by $Z_2$ to thus obtain
\begin{equation}\label{SDK}
 \begin{bmatrix} S(\lambda_i,\lambda_j) & {I}(\lambda_i,\lambda_j) \\
 D(\lambda_i,\lambda_j) & S(\lambda_j,\lambda_i) \end{bmatrix} =: K(\lambda_i,\lambda_j).
 \end{equation}
 Furthermore, let $K_J$ denote the corresponding $2 \times 2$ matrix integral operator with kernel $K(x,y)$ supported on $J$.
 Then we know from \cite[eq.~(9.181)]{Fo10} that (\ref{8.2}) and (\ref{8.3}) imply
 \begin{equation}\label{B3}
 ( E^{\rm rG}(J;\xi))^2 = \det ( \mathbb I_2 - \xi K_J ).
 \end{equation}
 Since
 \begin{equation}\label{B3a} 
 p_0(s) = {1 \over \rho} {d^2 \over d s^2} E((0,s);\xi = 1),
\end{equation}
to obtain the analogues of (\ref{10.1}) and (\ref{10.2}), it suffices to deduce the corresponding expansions of the Fredholm
determinant in (\ref{B3}). In fact the sought expansions --- but not derived from (\ref{B3}) --- are already in the literature, and reveal unobvious relations with other
point processes, as we will now proceed to detail.

\section{Relationship to diffusion processes}
\subsection{The coalescence process $A + A \to A$}
In recent years attention has focussed on (\ref{1}), and its analogue for the $k$-th nearest neighbour spacings distributions, as
an ansatz for the spacing distribution in certain one-dimensional non-equilibrium statistical mechanical models \cite{GT07,GT08}.
In fact, as will now be revised, (\ref{1}) is the exact result for the $t \to \infty$ rescaled consecutive spacing distribution in diffusion limited
coalescence $A + A \to A$.

Let us first define the latter. Starting from a point distribution on the real line with uniform density $\rho$,
each particle performs independent Brownian motion. When two particles meet they immediately merge into one.
The dynamical particle density $\rho(x;t)$ therefore decreases with time. One has $\rho(x;t) \propto t^{-1/2}$ 
\cite[Eq.~(16)]{bA98},
and in particular the system is not in equilibrium. Nonetheless, by rescaling the particle positions so that 
the density is again uniform and independent of time,  the $t \to \infty$ steady state has some special properties which we
will now revise. 

Specifically, consider the gap probability $E(0;J)$, $J = \{ (x_{2i-1}, x_{2i}) \}_{i=1,\dots,n}$,
with
 \begin{equation}\label{xx}
 x_1 < x_2 < \cdots <x_{2n-1} < x_{2n}.
\end{equation}
Then as $t \to \infty$, and with the density rescaled to a constant value $\rho$, one has the exact result \cite[Eq.~(11) with
eq.~(15) substituted]{bA98}
\begin{equation}\label{xa}
E^{\rm c}(0;J) = {\rm Pf} \, A
\end{equation}
where $A$ is the antisymmetric matrix with entries $(ij)$, $i < j$,
\begin{equation}\label{xb}
A_{ij} = {\rm erfc} \Big (  {\sqrt{\pi} \rho \over 2} (x_j - x_i) \Big )
\end{equation}
(the superscript $c$ on $E^{\rm c}$ refers to the coalescence process). 
One recalls that a Pfaffian of an even degree antisymmetric
matrix $A = [\alpha_{j,k}]_{j,k=1,\dots,2n}$, $\alpha_{j,k} = - \alpha_{k,j}$, is by definition
(see e.g.~\cite[Def.~6.1.4]{Fo10})
\begin{equation}\label{2.12a}
{\rm Pf} \, A = \sum_{P \in S_{2n} : P(2l) > P(2l-1)}  \varepsilon (P)
\alpha_{P(1),P(2)}  \alpha_{P(3),P(4)}  \cdots   \alpha_{P(2n-1),P(2n)} ,
\end{equation}
where $S_{2n}$ is the set of all permutations of $\{1,2,\dots,2n\}$, the restriction $P(2l) > P(2l-1)$ is required for
each $l=1,\dots,n$, and furthermore the sum is restricted to distinct terms only.
In \cite{bA98} the expression for (\ref{xa}) is the summation corresponding to the RHS of (\ref{2.12a}) only,
and in particular the fact that this is a Pfaffian is not mentioned.
In the simplest case of $J = (0,s)$, (\ref{xa}) and (\ref{xb}) give
\begin{equation}\label{2.12a}
E^{\rm c}(0;(0,s)) = {\rm erfc} \Big ( {\sqrt{\pi} \rho \over 2} s \Big )
\end{equation}
and application of (\ref{B3a}) implies
$$
p_0^{\rm c}(s) = {\rho \pi s \over 2} e^{-\pi (\rho s)^2/4},
$$
thus realizing the Wigner surmise (\ref{1}). 

One significance of (\ref{xa}), (\ref{xb}) is that the exact form of the $n$-point correlation then follows from the general formula
\begin{equation}\label{3.4}
\rho_{(n)}(y_1,\dots,y_n) = (-1)^n {\partial^n \over \partial x_2 \partial x_4 \dots \partial x_{2n}}
E(0; (x_1,x_2),\dots,(x_{2n-1},x_{2n})),
\end{equation}
with the RHS evaluated at $x_{2i} = x_{2i-1} = y_i$, $(i=1,\dots,n)$. Specifically, we read off from \cite[eq.~(20)]{MbA01} that
\begin{equation}\label{5.1}
\rho_{(n)}(y_1,\dots,y_n) = (-\rho)^n {\rm Pf} \, \tilde{A},
\end{equation}
where $\tilde{A}$ is the anti-symmetric matrix with upper triangular entries
\begin{align}\label{4.1}
 \tilde{A}_{2i-1,2j-1} & = {\rm erfc} \Big ( {\sqrt{\pi} \rho \over 2} (y_j - y_i) \Big ),    & \tilde{A}_{2i-1,2j}   = - e^{-\pi \rho^2 (y_j - y_i)^2 / 4} \nonumber \\
    \tilde{A}_{2i,2j-1}  & =  e^{-\pi \rho^2 (y_j - y_i)^2 / 4},    & \tilde{A}_{2i,2j}   = - {\pi \rho \over 2} (y_j - y_i)  e^{-\pi \rho^2 (y_j - y_i)^2 / 4}.
\end{align}
This assumes the ordering $y_1 < \cdots < y_n$. However, noting that for $y_j > y_i$
\begin{equation}\label{6.1}
{\rm erfc} \Big ( {\sqrt{\pi} \rho \over 2} (y_j - y_i) \Big )  = {\rm sgn} (y_j - y_i) - {2 \over \sqrt{\pi}}
\int_0^{\sqrt{\pi} \rho (y_j - y_i)/2} e^{-t^2} \, dt
\end{equation}
we see that $\tilde{A}_{2i-1,2j-1}$, like all other entries in (\ref{4.1}), is anti-symmetric with respect to the interchange $y_i \leftrightarrow y_j$,
and thus we can drop the ordering requirement since the Pfaffian in (\ref{5.1}) is then symmetric.
As with (\ref{xa}), the form of (\ref{5.1}) given in \cite{MbA01} is a summation corresponding
to the RHS of (\ref{2.12a}), without the identification of the antisymmetric matrix $\tilde{A}$ nor the Pfaffian.
The latter first appeared in 
\cite[Theorem 2]{TZ11}.
We remark too that in this reference the role of the initial condition
is discussed with more precision.

If we write
$$
S(x,y;\rho) = {\rho \over 2} e^{- \pi \rho^2 (x - y)^2/4}
$$
then the RHS of (\ref{6.1}) can be rewritten
\begin{equation}\label{8.2c}
{\rm sgn}(y_j - y_i) - 2 \int_{y_i}^{y_j} S(y_i,u;\rho) \, du =: 2 {I}(y_i,y_j;\rho)
\end{equation}
while
$$
  \tilde{A}_{2i,2j}  = {2 \over \rho^2} {\partial \over \partial x} S(x,y;\rho) \Big |_{x = y_i \atop y = y_j} =: {2 \over \rho^2} D(y_i,y_j).
$$
It follows that (\ref{5.1}) can be rewritten in the block form \cite[Theorem 2]{TZ11}
\begin{align}\label{8.1}
\rho_{(n)}^{\rm c}(y_1,\dots,y_n) = 2^n    {\rm Pf}
\begin{bmatrix} - {I}(\lambda_i,\lambda_j;\rho) & S(\lambda_i,\lambda_j;\rho) \\
- S(\lambda_j,\lambda_i;\rho) & D(\lambda_i,\lambda_j;\rho) \end{bmatrix}.
\end{align}
Comparison with (\ref{8.2}) shows
\begin{equation}\label{8.2q}
\rho_{(n)}^{\rm c}(y_1,\dots,y_n) \Big |_{\rho = \sqrt{2/\pi}} = 2^n \rho_{(n)}^{\rm rG} (y_1,\dots,y_n),
\end{equation}
a fact that can also be read off from \cite[Theorem 1 and Corollary 4]{TZ11}.
This has the interpretation that if every particle in the $t \to \infty$ coalescence process $A + A \to A$,  rescaled to have density
$\rho = \sqrt{2/\pi}$, is deleted with probability $1/2$ a point process identical to that of the bulk real eigenvalues of the real Ginibre ensemble
results (cf.~the sentence below (\ref{8.3})). Making use of (\ref{8.3}) it follows that
\begin{equation}
E^{\rm c}(0;J)  \Big |_{\rho = \sqrt{2/\pi}} = E^{\rm rG}(J;2).
\end{equation}

Recalling (\ref{xa}), and with $J = \{(x_{2i-1}, x_{2i}) \}$, we thus have
\begin{equation}
E^{\rm rG}(J;2) = \det \Big [ {\rm sgn}(x_j - x_i) {\rm erfc} \Big ({1 \over \sqrt{2}} |x_j - x_i| \Big ) \Big ],
\end{equation}
where use has been made of the classical identity $({\rm Pf} A)^2 = \det A$. According to the discussion below (\ref{8.3}) and with $\xi$ fixed, 
only for $0 < \xi \le 1$ does $E^{\rm rG}(J;\xi)$ directly relate to an observable quantity. On the other hand, it does provide a sum rule linking the
quantities $\{E^{\rm rG}(k;J) \}_{k=0,1,\dots}$.

To fulfill our aim of obtaining the analogue of the expansions  (\ref{10.1}) and (\ref{10.2}), we now turn our attention to a different diffusion process.

\subsection{The annihilation process $A + A \to \emptyset$}
In the annihilation process on the line particles are freely diffusing but with the condition that colliding particles annihilate. The rescaled $t \to \infty$
limit of this system can be analyzed exactly using methods that parallel those used to analyze the coalescence process $A + A \to A$.
But rather than $E(0;J)$ admitting a structured formula like (\ref{xa}), it is the quantity $E({\rm even};J)$ that the interval $J$ contains
an even number of particles that is structured. In fact \cite{MbA01}
\begin{equation}\label{3.7}
E^{\rm a}({\rm even};J) = {1 \over 2} + {1 \over 2} E^{\rm c}(0;J).
\end{equation}
The relation (\ref{3.7}) implies, via (\ref{3.4}) (which remains valid with $E(0;J)$ replaced by $E({\rm even};J)$), that 
 \begin{equation}\label{3.8}
 \rho^{\rm a}(x_1,\dots,x_n) = 2^{-n}   \rho^{\rm c}(x_1,\dots,x_n),
\end{equation}
which was known to a number of authors (see \cite{MbA01} and references therein).
Substituting in (\ref{8.2}) tells us that
  \begin{equation}\label{9.3}
 \rho^{\rm a}(y_1,\dots,y_n) \Big |_{\rho = 1/\sqrt{2 \pi}}=   \rho^{\rm rG}(y_1,\dots,y_n) .
\end{equation}
As already noted below (\ref{6.1}), the reference \cite{TZ11} contains the explicit Pfaffian expression (\ref{5.1}) and
(\ref{8.1}). It was observed by a referee that the latter, upon multiplication by $2^{-n}$ as required in (\ref{3.8}),
coincides with (\ref{8.2}), (\ref{8.2a}). Interestingly, the follow up paper  \cite{TZ12} shows that the dynamics of the annihilation
process, with a particular infinite density initial condition, is specified by an extended Pfaffian kernel. There is
also a natural dynamics for real Ginibre matrices --- choose the elements to now evolve according to
Dyson Brownian motion. It is not known how to fully characterise the dynamics, but the exact dynamical
two-point function is known, and this is distinct from the corresponding quantity for the annihilation process
\cite{TZ14}.

An immediate consequence of (\ref{3.8}), obtained by substituting (\ref{9.3}) in (\ref{8.3}), is that
\begin{equation}\label{si}
E^{\rm a}(J;\xi) \Big |_{\rho = 1/\sqrt{2 \pi}} = E^{\rm rG}(J;\xi),
\end{equation}
and thus in particular that
\begin{equation}\label{Eh}
E^{\rm a}(0;J)  \Big |_{\rho = 1/\sqrt{2 \pi}} = E^{\rm rG}(0;J).
\end{equation}
This is significant, since $E^{\rm a}(0;(0,s))$ has been studied as the continuum limit of the zero temperature Glauber dynamics of the
one-dimensional Ising model, and this has led to the exact result \cite{DZ96}
\begin{equation}\label{9.4}
E^{\rm a}(0;(0,s)) \Big |_{\rho = 1} =
\Big ( \sqrt{1 - \mu A_1(s)} - (q - 1) \sqrt{- \mu A_1(s)} \Big ) e^{A_2(s)},
\end{equation}
where
\begin{align}
A_1(z) & = \sum_{n=1}^\infty (-2 \mu)^n \int_0^z dx_1 \cdots  \int_0^z dx_n \, g(z,x_1) {\partial \over \partial x_1} g(x_1,x_2) \cdots 
 {\partial \over \partial x_1} g(x_n,z) \label{9.4a} \\
A_2(z) & =  - {1 \over 2} \sum_{n=1}^\infty {(-2 \mu)^n  \over n}
\int_0^z dx_1 \cdots  \int_0^z dx_n \,  {\partial \over \partial x_1} g(x_1,x_2) \cdots 
 {\partial \over \partial x_1} g(x_n,x_1), \label{9.4b}
 \end{align}
 with
 $$
 g(x,y) = {2 \over \sqrt{\pi}} \int_0^{\sqrt{\pi}(y-x)q/(2(q-1))} e^{-u^2} \, du, \quad \mu = {q - 1 \over q^2}, \quad q = 2.
 $$

 Moreover, (\ref{9.4}) has been used to compute both the small $s$ expansion
 \begin{align}\label{Es}
p^{\rm a}(0;s) = & \pi  s-\pi ^2 s^3+\frac{\pi ^2 s^4}{3}+\frac{\pi ^3 s^5}{2}-\frac{4 \pi ^3 s^6}{15}-\frac{\pi ^4 s^7}{6}+\frac{7 \pi ^4 s^8}{60}+\frac{\pi ^5 s^9}{24}\nonumber \\
& - \frac{23 \pi ^5 s^{10}}{630}-\frac{\pi ^6 s^{11}}{120}+\frac{1523 \pi ^6 s^{12}}{166320}+\frac{\pi ^6 (-64+2520 \pi ) s^{13}}{1814400}+O(s^{14})
\end{align}
and the large $s$ expansion
 \begin{equation}\label{Esu}
 E^{\rm a}(0;(0,s)) = e^{- c_1 s + c_2 + o(1)}
 \end{equation}
 respectively, where in (\ref{Esu})
 \begin{align}
 c_1 & = {1 \over 2} \zeta(3/2) \approx 1.3062 \label{Es1} \\
 c_2 & = \log 2 - {1 \over 4} + {1 \over 4 \pi} \sum_{n=2}^\infty {1 \over n} \Big ( - \pi + \sum_{p=1}^{n-1} {1 \over \sqrt{p(n-p)}} \Big ) \approx 0.0627. \label{Es2}
 \end{align}   
 A crucial aspect of (\ref{Esu}) is that it relies on the large $s$ expansion of (\ref{9.4}) begin computed first for general $0< q < 2$, and then taking the
 limit $q \to 2$. 
 Hence (\ref{Esu}) is not a rigorous statement, although in the paragraph including (\ref{dl}) below the leading
 term will be established rigorously.
 The necessity of such a limiting procedure is immediately apparent from the large $s$ form of the first factor in (\ref{9.4})
 \cite{DHP96}
 $$
 \Big ( \sqrt{1 - \mu A_1(s)} - (q - 1) \sqrt{- \mu A_1(s)} \Big ) = \sqrt{q(2-q)} + o(1).
 $$
 The derivation of this results makes use of the relationship between $A_1(s)$ and the elements of a certain matrix inverse.
 
 The derivation of the asymptotic formula for $A_2(s)$ makes use of the fundamental asymptotic expansion of the Wiener-Hopf determinant of
 an integral operator with difference kernel $K(x,y) = K(x-y,0)$ defined on $(0,s)$ for $s \to \infty$ \cite{Ka54},
 \begin{align}\label{sq1}
 &  \log \det ( \mathbb I - \xi K_{(0,s)}) \nonumber \\
 & = {s \over 2 \pi} \int_{-\infty}^\infty \log (1 - \xi \tilde{K}(u)) \, du +
{1 \over 4 \pi^2}  \int_0^\infty u \Big | \int_{-\infty}^\infty e^{-iku} \log (1 - \xi \tilde{K}(k)) \, dk \Big |^2 \, du + o(1),
\end{align}
where $\tilde{K}(u) = \int_{-\infty}^\infty e^{i x u} K( x,0) \, dx$. This is used in conjunction with the identity (see e.g.~\cite[Exercises 9.3 q.1]{Fo10}))
\begin{equation}\label{Rr}
\log \det ( \mathbb I - \lambda K_{(0,s)}) = - \sum_{p=1}^\infty {\lambda^p \over p}
\int_0^s dx_1 \cdots \int_0^s dx_p \, \prod_{l=1}^p K(x_l,x_{l+1})
\end{equation}
where $x_{p+1} := x_1$.

The sought analogue of the expansions  (\ref{10.1}) and (\ref{10.2}) for the gap probability of the real eigenvalues in the real Ginibre ensemble
is an immediate corollary of (\ref{Eh}), (\ref{Es1}) and (\ref{Es2}).

\begin{cor}\label{C1}
The gap probability of the real eigenvalues in the bulk of real Ginibre ensemble, which have $\rho = 1/\sqrt{2 \pi}$, has the small $s$ expansion
\begin{align}\label{Rr1}
& E^{\rm rG}(0;(0,s)) = 1-\frac{s}{\sqrt{2 \pi }}+\frac{s^3}{12 \sqrt{2 \pi }} -\frac{s^5}{80 \sqrt{2 \pi }}+\frac{s^6}{720 \pi }  \nonumber \\
&+\frac{s^7}{672 \sqrt{2 \pi }}-\frac{s^8}{3360 \pi }-\frac{s^9}{6912 \sqrt{2 \pi }}+\frac{7 s^{10}}{172800 \pi } 
 +\frac{s^{11}}{84480 \sqrt{2 \pi }} -\frac{23 s^{12}}{5322240 \pi } \nonumber \\
&-\frac{s^{13}}{1198080 \sqrt{2 \pi }}+\frac{1523 s^{14}}{3874590720 \pi }-
\Big ( \frac{1}{762048000 \sqrt{2} \pi ^{3/2}}+\frac{1}{19353600 \sqrt{2 \pi }} \Big )s^{15} + O(s^{16})
\end{align}
and the large $s$ expansion
\begin{equation}\label{Rr2}
E^{\rm rG}(0;(0,s)) = e^{- \tilde{c}_1 s + c_2 + o(1)},
\end{equation}
where $c_2$ is as in (\ref{Es2}) while
\begin{equation}\label{Rr3}
\tilde{c}_1 = {1 \over \sqrt{2 \pi}} c_1 = {1 \over 2 \sqrt{2 \pi}} \zeta(3/2).
\end{equation}
\end{cor}

We remark that the small $s$ expansion of $p^{\rm rG}(0;(0,s))$ follows from (\ref{Rr1}) by the formula (\ref{B3a}); for us it is more convenient to consider
$ E^{\rm rG}(0;(0,s))$ as this is more readily accessible numerically, as we will soon demonstrate. But before doing so, we make note of a second
corollary of the exact results for the annihilation process, in particular (\ref{3.7}) and (\ref{xa}) and as they apply to real Ginibre ensemble.

\begin{cor}\label{C2}
Let $A$ be the antisymmetric matrix specified by (\ref{xb}). We have
\begin{equation}\label{60}
E^{\rm rG}({\rm even};J) = {\rm Pf} \, A \Big |_{\rho = \sqrt{2/ \pi}}
\end{equation}
and in particular
\begin{equation}\label{61}
E^{\rm rG}({\rm even};(0,s)) = {1 \over 2} + {1 \over 2} {\rm erfc} {s \over \sqrt{2}}.
\end{equation}
\end{cor}

\section{Comparison with numerical computations}
We have two formulas for $E^{\rm rG}(0;(0,s))$ --- the one implied by setting $\xi = 1$ in (\ref{B3}), and the one that follows by replacing $s$ by $s/\sqrt{2 \pi}$
in (\ref{9.4}). It has already been remarked that many Fredholm determinant formulas in random matrix theory are well suited to high precision numerical
computation. However this requires that the kernel be analytic, whereas (\ref{8.2c}) shows that the entry ${I}(x,y)$ is not differentiable at $x=y$, and
thus disallowing this approach. In relation to (\ref{9.4}), according to (\ref{Rr}), (\ref{9.4b}) can be expressed in terms of a Fredholm determinant, and
furthermore the corresponding kernel is analytic. But the quantity (\ref{9.4a}) does not have a structure consistent with (\ref{Rr}), and so again we run into difficulties.

Nonetheless, it is still possible to illustrate the validity of Corollary \ref{C1}. This can be done by noting an $N/2 \times N/2$  (for convenience it will
be assumed $N$ is even) determinant formula for
$E^{{\rm rG},N}(0;(-s,s))$ in the finite $N$ real Ginibre ensemble, and then evaluating this numerically for large $N$.

For this purpose, with $k$, $N$ even let $p(\{\lambda_l\}_{l=1,\dots,k}, \{x_j \pm i y_j \}_{j=1,\dots,(N-k)/2})$ denote the probability density that there are
$k$ real eigenvalues and $N-k$ complex eigenvalues at the prescribed locations. Define the generalized partition function
\begin{align*}
& Z_N[u,v] \nonumber \\
& \quad = \sum_{k=0 \atop k \: {\rm even}}^N 
\int_{\mathbb R} d \lambda_1 \cdots d \lambda_k \, \prod_{j=1}^k u(\lambda_l) \int_{\mathbb R_+^2} dx_1 dy_1 \cdots
 \int_{\mathbb R_+^2} dx_{(N-k)/2} dy_{(N-k)/2} \, \prod_{l=1}^{(N-k)/2} v(x_l,y_l).
 \end{align*}
 We see from this that
 \begin{equation}\label{ZE}
 Z_N[1 - \chi_{\lambda \in (-s,s)}, 1] = E^{{\rm rG}, N}(0;(-s,s))
 \end{equation}
 where $\chi_A = 1$ for $A$ true and $\chi_A = 0$ otherwise. 
 The significance of (\ref{ZE}) is that the generalized partition function admits the Pfaffian form \cite{Si06,FN07}
  \begin{equation}\label{ZE1}
 Z_N[u,v]= {2^{-N(N+1)/4} \over \prod_{l=1}^N \Gamma(l/2)} {\rm Pf} \, [ \alpha_{j,l}[u]+ \beta_{j,l}[v] ]_{j,l=1,\dots,N},
 \end{equation}
where, with $\{p_j(x)\}_{j=0,1,\dots}$ monic polynomials of the labelled degree but otherwise arbitrary,
  \begin{equation}\label{13.0}
\alpha_{j,l}[u]= \int_{-\infty}^\infty dx \, u(x)   \int_{-\infty}^\infty dy \, u(y) e^{-(x^2+y^2)/2} p_{j-1}(x)   p_{l-1}(x) {\rm sgn} (y-x)
 \end{equation}
 (the explicit form of $\beta_{j,l}$ is also known, but this will not be required below). Moreover, if the monic polynomials are
 chosen as
 \begin{equation}\label{13.1}
 p_{2j}(x) = x^{2j}, \qquad p_{2j+1}(x) = x^{2j+1} - 2j x^{2j-1},
 \end{equation}
 then for $u=v=1$ and with $p<q$ the skew orthogonality
  \begin{equation}\label{13.2} 
  \alpha_{p,q}[1]+\beta_{p,q}[1] =
  \left \{ \begin{array}{ll} r_{j-1} := 2 \sqrt{2 \pi} \Gamma(2j-1), & (p,q) = (2j-1,2j) \\
  0, & {\rm otherwise}, \end{array} \right.
 \end{equation}
 holds true \cite{FN07}.
 
 The above theory can be used to give the sought   determinant formula for $E^{{\rm rG},N}(0;(-s,s))$ can be derived.
 
 \begin{prop}
 With $\gamma(a;x) = \int_0^x t^{a-1} e^{-t} \, dt$ denoting the (lower) incomplete gamma function, we have
 \begin{equation}\label{14.1}
E^{{\rm rG},N}(0;(-s,s)) = \det \Big [ \delta_{j,l} - {\gamma(l+j-3/2;s^2) \over \sqrt{2 \pi} (\Gamma(2j-1) \Gamma(2l-1))^{1/2}}
\Big ]_{j,l=1,\dots,N/2}.
\end{equation}
\end{prop}

\noindent
Proof. \quad With the choice (\ref{13.1}) it follows that for $l> j$
$$
\alpha_{j,l}[u]+ \beta_{j,l}[1] = \alpha_{j,l}[u] -   \alpha_{j,l}[1] + 
 \left \{ \begin{array}{ll} r_{(j-1)/2}, & j \: {\rm odd}, \: l = j+1 \\
  0, & {\rm otherwise}. \end{array} \right.
$$
Furthermore, for $u(x)$ even we have $\alpha_{j,l}=0$ for $j,l$ both of the same parity. This means that every second element in the Pfaffian vanishes
(i.e.~there is a checkerboard pattern of zero entries --- this same feature has previously been noted in relation to the computation of their
being exactly $k$ real eigenvalues \cite{AK07,FN08p,Ma11}). In this circumstance it is generally true that
$$
{\rm Pf} \, [c_{j,l} ]_{j,l=1,\dots,N} = \det [ c_{2j-1,2l}]_{j,l=1,\dots,N/2},
$$
and so we have
 \begin{equation}\label{14.2}
 Z_N[u,v=1] = {2^{-N(N+1)/4} \over \prod_{l=1}^N \Gamma(l/2)}
 \det \Big [ \delta_{j,l} r_{j-1} + \alpha_{2j-1,2l}[u] -   \alpha_{2j-1,2l}[1] \Big ]_{j,l=1,\dots,N/2}.
 \end{equation} 
 
 Now we set $u = 1 - \chi_{\lambda \in (-s,s)}$. Substituting in (\ref{13.0}) we have
 \begin{align}\label{15.1}
   \alpha_{2j-1,2l}[u] -   \alpha_{2j-1,2l}[1] = & 
   \Big ( \Big ( \int_{-\infty}^\infty - \int_{-s}^s \Big ) dx   \Big ( \int_{-\infty}^\infty - \int_{-s}^s \Big ) dy -
  \int_{-\infty}^\infty dx      \int_{-\infty}^\infty dy    \Big ) \nonumber \\
  & \times e^{-(x^2 + y^2)/2} p_{2j-2}(x) p_{2l-1}(y) {\rm sgn}(y-x).
\end{align}
Noting from (\ref{13.1}) that
$$
p_{2l-1}(x) = - e^{x^2/2} {d \over dx} \Big ( e^{-x^2/2} p_{2l-2}(x) \Big )
$$
allows (\ref{15.1}) to be simplified to read
$$
 \alpha_{2j-1,2l}[u] -   \alpha_{2j-1,2l}[1] = - 2 \int_{-s}^s e^{-y^2} p_{2l-2}(y) p_{2j-2}(y) \, dy = - 2
 \gamma(l+j-3/2;s^2).
 $$
Substituting this result in (\ref{14.2}) and performing some simple manipulations we arrive at (\ref{14.1}). \hfill $\square$

\medskip
We have used (\ref{14.1}) to compute $E^{{\rm rG},N}(0;(-s/2,s/2))$ with $N = 120$, and compared it against the expansions (\ref{Rr1}) and (\ref{Rr2}).
Graphical accuracy in the case of the small $s$ expansion (\ref{Rr1}) is found for $s < 2.2$;  forming $(\log E^{{\rm rG},N}(0;(-s/2,s/2)))/s$ graphical
accuracy is obtained in the case of the large $s$ expansion (\ref{Rr2}) for $s > 1.3$.  Thus the expansions in Corollary \ref{C1} are indeed consistent
with numerical computations.

\section{Concluding remarks}
Our approach to deducing the small and large $s$ expansions in Corollary 1\ref{C1} relies on the equivalence in distribution of the 
bulk scaling limit  of the real eigenvalues for the real Ginibre ensemble and the
rescaled $t \to \infty$ limit of the annihilation process $A + A \to \emptyset$. It is thus of interest to remark that certain diffusions also play a
crucial role in the large distance asymptotic analysis of the gap probability for so called real Weyl random polynomials
\cite{SM07,SM08} (see also \cite{DPSZ02}). These are random polynomials $p_N(z) = \sum_{n=0}^N a_n {z^n \over \sqrt{n!}}$ where the $a_n$ are
independent real standard Gaussians. For large $N$ the density of real zeros on the real axis is to leading order equal to $1/\pi$. It turns
out that the probability of there being no real zeros on the interval $(-s,s)$ can, for large $s$, be related to the probability that a scalar Gaussian
random field, with a certain covariance, and subject to random initial conditions, at any chosen point keeps the same sign it had initially. This statement
requires that the diffusion occurs in the limit of large dimension $d$, and leads to the prediction that $E(0,(-s,s))$ has the leading
large $s$ form $e^{-2 \theta_\infty s}$ where $\theta_\infty$ is a certain so called persistence exponent. It's exact value is not known, although the numerical
estimate $\theta_\infty \approx 0.41$ is quoted in \cite{SM08}. Note in particular that the decay is as an exponential as in (\ref{Rr2}) and not a Gaussian
as in the Wigner surmise (\ref{1}). Generally an exponential form will result when the point process is compressible and thus
$\int_{-\infty}^\infty( \rho_{(2)}^T(x,0)/\rho +  \delta(x)) \, dx >0$ \cite{FP92}; for the real Ginibre eigenvalues this quantity equals $2 - \sqrt{2}$ \cite{FN07}.

Suppose we didn't have available the small and large distance expansions (\ref{Es}) and (\ref{Esu}) for the annihilation process. Could we still
derive the expansions of Corollary \ref{C1}? Thus we are asking if the results of Corollary \ref{C1} can be derived using (\ref{B3}) or (\ref{8.3}) and (\ref{8.2}).
Certainly the small distance expansion (\ref{Rr1}) can in principle be derived by expanding the entries (\ref{8.2a}) for small argument,
and then substituting in (\ref{8.2}) and expanding out the Pfaffian
to expand the $\rho_{(n)}$ for small argument, before computing the integrals as required in (\ref{B3}). The explicit form of the joint
eigenvalue probability density function $p(\{\lambda_l\}_{l=1,\dots,k}, \{x_j \pm i y_j \}_{j=1,\dots,(N-k)/2})$ for the real and complex
eigenvalues \cite{Ed95} tells us that for small distances
$\rho_{(n)}(\lambda_1,\dots,\lambda_n)$ is proportional to $\prod_{1 \le j < k \le n} |\lambda_k - \lambda_j|$ and thus for small $s$
$\rho_{(n)}$ first contributes to (\ref{8.3})  at order $s^{n(n+1)/2}$. Therefore to obtain the expansion (\ref{Rr1}), which is accurate up to the term
proportional to $s^{15}$, the series (\ref{8.3}) can be truncated at $n=4$. An analogous approach could also be used to compute the small
distance expansions of $E^{{\rm rG}}(k;(0,s))$ for $k>0$, although for this to be practical $k$ itself must be small.

In relation to the large distance expansion (\ref{Rr2}), we require a matrix kernel analogue of (\ref{sq1}). This is known (see e.g.~\cite[pg.~14]{BW04}),
but it seems that only the leading term is readily computable and simply related to the corresponding term in (\ref{sq1}).
Thus we have that for $s \to \infty$
\begin{equation}\label{dl}
 \log \det( \mathbb I_2 - \xi K_{(0,s)}) = {s \over 2 \pi} \int_{-\infty}^\infty \log \det (\mathbb I_2 - \xi \tilde{K}(u)) \, du + O(1),
\end{equation}
where $\tilde{K}(u)$ refers to the component wise Fourier transform of the kernel corresponding to $K_{(0,s)}$.
Taking the Fourier transform of the matrix elements (\ref{SDK}) shows
$$
\tilde{S}(k) = e^{-k^2/2}, \qquad \tilde{D}(k) = i k e^{-k^2/2}, \qquad \tilde{I}(k) = {1 \over ik} \Big ( -1 + e^{-k^2/2} \Big )
$$
Thus
$$
\det (\mathbb I_2 - \xi \tilde{K}(u)) = 1 - (2 \xi - \xi^2) e^{-k^2/2}
$$
Substituting in (\ref{dl}) and setting $\xi = 1$ allows the integral to be evaluated, and the leading term in (\ref{Rr2}) is reclaimed. 
Moreover, this provides a rigorous derivation of (\ref{Rr2}) to leading order.
The computation
of the large distance expansion of $E^{{\rm rG}}(k;(0,s))$ for $k>0$ remains an open problem (see the recent review \cite{Fo12e} for methods used
in the analogous quantity for the bulk scaled GOE).

Lastly, we will comment on a number of problems relating to the present one which require more investigation. Already noted is the task of
providing high precision computation of $E^{{\rm rG}}(0;(0,s))$, or more generally $E^{{\rm rG}}(0;(0,s))$. Another is to obtain the asymptotics
not of the gap probability but rather the hole probability. Thus the eigenvalues of real Ginibre matrices are in general complex, so a natural
quantity is the hole probability that there are no eigenvalues in a disk of circumference $\alpha$ centred at the origin of the complex plane.
In the case of the complex Ginibre ensemble (Gaussian random matrices with independent complex Gaussian entries) the large $\alpha$
asymptotics have been computed in \cite{Fo92c} (see also \cite{APS09}), and in the case of complex Weyl polynomials the leading
asymptotic form is also known \cite{ST05,Ni10}. Finally, we draw attention to the task of computing the asymptotics of the gap
probability for the real eigenvalues of the real Ginibre ensemble scaled not in the bulk, but at the edge. The formula (\ref{B3}) still holds,
but now the entries of the kernel (\ref{8.2a}) require that $S(x,y)$ be replaced by \cite{FN07}
$$
S(x,y) = {1 \over \sqrt{2\pi}} \Big ( {1 \over 2} e^{-(x-y)^2/2}
\Big ( 1 - {\rm erf} {x + y \over \sqrt{2}} \Big ) + {e^{-y^2} \over 2 \sqrt{2}}(1 + {\rm erf} \, x) \Big ).
$$

\medskip
\noindent
{\it Note added:} This paper was written in 2013 while the author was on long service leave, and no publication sought until the
announcement of the JPhysA special issue in honour of his PhD supervisor, Professor R.J.~Baxter. Very recently the
result (\ref{Esu}) has been used in the work of \cite{KPTTZ15} on a study of the asymptotic form of probability that an $N \times N$
real Ginibre matrix has a fixed number $k$ of real eigenvalues in the limit $N \to \infty$.

\subsection*{Acknowledgements}
Financial support for this work from the Australian Research Council is acknowledged. I thank Carlo Beenakker for sending me a copy
of \cite{BEDPSW13} and correspondence which pointed out the numerical studies therein of the spacing distribution for a certain
ensemble of matrices related to the real Ginibre ensemble. I also thank Jonathan Edge for going to the trouble of comparing the
asymptotic formula of this work against large scale simulation data extending \cite{BEDPSW13}, and by so doing identifying an error in the 
reporting of the value
of $c_2$ in the original version of this paper.

% \bibliographystyle{amsplain}
%\bibliography{book1}

\providecommand{\bysame}{\leavevmode\hbox to3em{\hrulefill}\thinspace}
\providecommand{\MR}{\relax\ifhmode\unskip\space\fi MR }
\providecommand{\MRhref}[2]{%
  \href{http://www.ams.org/mathscinet-getitem?mr=#1}{#2}
}
\providecommand{\href}[2]{#2}

\end{document}